\documentclass[aip,jcp,10pt,floatfix,twocolumn]{revtex4-2} 
\usepackage{color,soul} 
\usepackage{float}
\usepackage{graphicx}
\usepackage{amsmath, amsthm, amssymb, mathtools}
\usepackage{multirow} 
\usepackage[normalem]{ulem}
\def\be{\begin{equation}}
\def\ee{\end{equation}}
\begin{document}

\title{The conformational phase diagram of charged polymers in the presence of attractive bridging crowders} 
\author{Kamal Tripathi} 
\email{kamalt@imsc.res.in}
\affiliation{The Institute of Mathematical Sciences, C.I.T. Campus, Taramani, Chennai 600113, India} 
\affiliation{Homi Bhabha National Institute, Training School Complex, Anushakti Nagar, Mumbai 400094, India}
\affiliation{Univ. Grenoble Alpes, CNRS, Grenoble INP, 3SR, F-38000 Grenoble, France} 
\author{Hitesh Garg} 
\email{hiteshgarg@imsc.res.in}
\affiliation{The Institute of Mathematical Sciences, C.I.T. Campus, Taramani, Chennai 600113, India} 
\affiliation{Homi Bhabha National Institute, Training School Complex, Anushakti Nagar, Mumbai 400094, India}
\author{R. Rajesh} 
\email{rrajesh@imsc.res.in}
\affiliation{The Institute of Mathematical Sciences, C.I.T. Campus, Taramani, Chennai 600113, India} 
\affiliation{Homi Bhabha National Institute, Training School Complex, Anushakti Nagar, Mumbai 400094, India}
\author{Satyavani Vemparala} 
\email{vani@imsc.res.in} 
\affiliation{The Institute of Mathematical Sciences, C.I.T. Campus, Taramani, Chennai 600113, India} 
\affiliation{Homi Bhabha National Institute, Training School Complex, Anushakti Nagar, Mumbai 400094, India}

\begin{abstract}
Using extensive molecular dynamics simulations, we obtain the conformational phase diagram of a charged polymer in the presence of oppositely charged counterions and neutral attractive crowders for monovalent, divalent and trivalent counterion valencies. We demonstrate that the charged polymer can exist in three phases: (1) an extended phase for low charge densities and weak polymer-crowder attractive interactions ($CE$), (2) a collapsed phase for high charge densities and weak polymer-crowder attractive interactions, primarily driven by counterion condensation ($CCI$), and (3) a collapsed phase for strong polymer-crowder attractive interactions, irrespective of the charge density, driven by crowders acting as bridges or crosslinks ($CCB$). Importantly, the simulations reveal that the interaction with crowders can induce collapse, despite the presence of strong repulsive electrostatic interactions, and can replace condensed counterions to facilitate a direct transition from the $CCI$ and $CE$ phases to the $CCB$ phase. 
\end{abstract}
\maketitle

\section{Introduction} 

Charged polymers or polyelectrolytes (PEs), are ubiquitous in nature, encompassing biological examples like DNA and proteins, as well as synthetic counterparts such as block copolymers and sodium polystyrene sulfonate etc.~\cite{Wong-Rev,delaCruz2000, VirusRNA2006, VirusElectro2012, VirusGrossberg2016, PhysRevE.71.061928, BIP:BIP360311305, BLOOMFIELD1996334, mitragotri2015accelerating}. The conformations of highly charged PEs are influenced by a combination of factors, including long-range electrostatic interactions, counterion entropy, solvent quality, temperature, counterion valency, and dielectric constant of the solvent, which are reasonably well-understood~\cite{russell2002rapid, buchmueller2000collapsed, murthy2000counterion, dias2003modeling,chauhan2008tertiary}.  With increase in PE backbone charge density, the oppositely charged counterions begin to condense onto the PE via Manning condensation~\cite{manning1969limiting, manning2018condensation}. Beyond a critical charge density, the PE undergoes a transition from an extended to a collapsed conformation, irrespective of solvent quality~\cite{varghese2011phase,brilliantov98, tom2016mechanism, tom2017regimes, Kremer1993, Kremer1995,winkler98, brilliantov98, Melnikov1999, Holm, anoop11, Dobrynin2005, Chertovich2016, Chertovich2016a}. Several mechanisms, such as ionic solid theory, dipole theory, and counterion fluctuation theory, have been proposed to explain various aspects of the counterintuitive collapse behavior of like-charged PEs ~\cite{brilliantov98, tom2016mechanism,tom2017regimes, bloomfield1997dna, cherstvy2010collapse, gordievskaya2018interplay, muthukumar2004theory, solis2000collapse,brilliantov98, Dobrynin2005, varghese2011phase, khokhlov1994, Kremer1993, Kremer1995, winkler98, brilliantov98,BoJon99, Holm,Chertovich2016, manning1969limiting, Khokhlov1980, khokhlov1994,pincus98}.

While the conformations of an isolated PE have been studied in detail, it is important to consider that PEs often exist within complex and crowded heterogeneous environments~\cite{rivas2004life, zhou2008macromolecular, rivas2016macromolecular, zimmerman1991estimation, ellis2003join, ellis2001macromolecular, zimmerman1993macromolecular}. In such environments, the conformational landscape of even neutral polymers is significantly altered by the presence of crowder molecules, which can induce novel phases such as bridging and depletion-induced collapsed phases (see details below) through specific or non-specific interactions~\cite{nayar2020small, zangi2009urea, mardoum2018crowding,nakano2017model, shin2015kinetics, liu2020non, taylor2020effects, asakura1954interaction, asakura1958interaction}. It is, thus, reasonable to expect that crowders will also affect the conformations of a PE, as the relative strength of the polymer-crowder interactions and electrostatic interactions is varied. Although few prior studies~\cite{nayar2023molecular,mukherjee2019osmolyte} have shown that the inclusion of crowders molecules stabilize the crowder-free equilibrium collapsed conformations of small charged or overall charge-neutral polymers, the specific mechanisms underlying crowder-polymer interactions and the intricate interplay between counterions and crowders remain unexplored. Thus, despite the widespread applications, a comprehensive understanding of how both crowders and counterions influence PE conformations is currently lacking and constitutes the central focus of this paper. 

Unlike the case of PEs, the effect of crowders on the conformations of neutral polymers is better studied. When the polymer-crowder interactions are purely repulsive, entropic effects can induce effective depletion induced attractive interactions between the monomers which in turn drive a collapse transition~\cite{zhou2008macromolecular, asakura1954interaction, asakura1958interaction, tripathi2019confined, tripathi2023conformational, miyazaki2022asakura, bhat1992steric, kang2015effects}. The depletion effects are enhanced with further attractive interaction among the crowders~\cite{garg2023conformational}. When the interactions between the polymer and the crowders exhibit weak attraction, the crowders assume the role of a typical good solvent, leading to a preference for extended polymer conformations. However, on further increase of polymer-crowder attractive interactions, a counterintuitive collapse transition (effective poor solvent condition) is observed even for a self-avoiding polymer where the crowders act as bridges or crosslinks between monomers.~\cite{antypov2008computer, heyda2013rationalizing, rodriguez2015mechanism, huang2021chain, garg2023conformational}. Recently, we obtained the detailed configurational phase diagram of neutral polymers in the presence of attractive crowders, by delineating phase transitions from extended conformations to collapsed conformations induced by both intra-polymer attraction and bridging effects due to crowders~\cite{garg2023conformational}.

The effect of crowders on the conformation of PEs is also of considerable interest in the context of membrane less biomolecular condensates~\cite{erdel2018formation, king2021phase, ryu2021current, azaldegui2021emergence}. These condensates form due to either polymer polymer phase separation (PPPS) or liquid liquid phase separation (LLPS), both driven by strong attractive interaction with crowders. Studied primarily from the perspective of chromatin organization, these two mechanisms of phase separation mainly differ in the relative strengths of chromatin-chromatin interactions and chromatin-crowder interactions. PPPS is driven by formation of bridges or cross links between distant parts along chromatin mediated by crowder molecules, thus generating a collapsed globule conformation. The bridging crowders do not undergo a phase separation in the absence of attractive interaction with chromatin. LLPS, on the other hand, leads to condensates resembling liquid-like crowder-chromatin droplets. Unlike PPPS, the crowders phase separate even in the absence of  chromatin. There are experimental and theoretical studies investigating the role of bridging interactions in chromatin reorganization via PPPS or LLPS~\cite{ryu2021bridging, brackley2020polymer, brackley2013nonspecific, barbieri2012complexity, murugesapillai2014dna, machida2018structural, malhotra2021unfolding, brackley2020bridging, biedzinski2020beyond}. However, despite chromatin being inherently a charged polymer, it is usually modeled as a coarse-grained neutral polymer in the presence of neutral crowders, thus ignoring the presence of charged counterions and long range interactions~\cite{ryu2021bridging, brackley2020polymer, brackley2013nonspecific, barbieri2012complexity, malhotra2021unfolding}. A study of PE in a crowded environment, provides a more realistic model for biopolymers such as chromatin and also makes it possible to investigate phenomena such as PPPS arising due to competing interactions with counterions and crowders. 

In this paper, we investigate the impact of attractive neutral crowders on the conformations of a single, similarly charged  PE in the presence of oppositely charged counterions. We employ extensive molecular dynamics simulations utilizing generic coarse-grained bead-spring models. We identify three distinct phases: (1) the $CCI$ phase, which corresponds to a collapsed state primarily driven by the condensation of counterions; (2) the $CE$ phase, characterized by an extended conformation of the polymer due to repulsive electrostatic interactions within the PE; and (3) the $CCB$ phase, representing a secondary collapsed state primarily induced by the presence of attractive bridging crowders, with a substantial number of crowders located within the collapsed conformation. By identifying the phase transitions among these phases, we obtain the phase diagram in the PE charge density and polymer-crowder interaction phase space for monovalent, divalent, and trivalent counterions.

\section{Model and methods \label{sec:model}} 

We model a flexible PE chain using a coarse-grained linear bead-spring model consisting of $N_m$ monomers of charge $-q$ (measured in units of $e$) linked together by harmonic springs with bond interaction energy
\begin{equation}
U_{bond}(r)=\frac{1}{2} k(r-r_0)^2,
\end{equation}
where $k$ is the spring constant, $r_0$ is the equilibrium bond-length and $r$ is the distance between two bonded monomers. We set $r_0=2^{1/6} \sigma$, $k=500 k_B T \sigma^{-2}$, where $\sigma$ is the size of the monomer in Lennard Jones (LJ) units (see below), $k_B$ is the Boltzmann's constant and $T$ is the temperature. $N_m/Z$ neutralizing counterions with charge $+Zq$, each of the same size and mass as the monomer, are added to the system, where $Z$ is the valency of the counterion ($Z = 1, 2, 3$). In addition, $N_c$ neutral crowders, of same size and mass as the monomer, are added such that the number density of the crowders is $\rho_c=N_c/L^3=0.05 \sigma^{-3}$,  where $L$ is the size of the simulation box. The monomer density is fixed at $N_m/L^{3} = 1.112 \times 10^{-3} \sigma^{-3}$.  For this box length,  the PE chain in its extended phase does not interact with its own periodic images.

Non-bonded pairs of particles interact via excluded volume interactions (PE monomers, counterions, crowders) as well as long ranged electrostatic interactions (PE monomers, counterions). The excluded volume interactions are modeled by the truncated and shifted LJ potential,
\begin{equation}
U_{jk}(r) =
4 \epsilon_{jk}\left[ \left(\dfrac{\sigma}{r}\right)^{12} - \left(\dfrac{\sigma}{r}\right)^{6} \right], ~r <  r^{jk}_c,
\label{eq:ShiftedTruncatedLJPot}
\end{equation}
where indices $j$ and $k$ can be $m$ (monomer), $i$ (counterion) or $c$ (crowder), and $r$ is the distance between two particles. The parameter $\epsilon_{jk}$ and cutoff $r^{jk}_c$ are pair-dependent while $\sigma$ is same for all particles.  The LJ parameters for the different pairs are listed in Table~\ref{tab:LJpar}. Other than  $m$--$c$ and  $c$--$c$ pairs, all other pair interactions are purely repulsive. In particular, this choice mimics an implicit good solvent condition for the polymer.
\begin{table}
\caption{\label{tab:LJpar} Parameters for the LJ potential [see Eq.~(\ref{eq:ShiftedTruncatedLJPot})] between different pairs of particles. The notations $m$, $i$ and $c$ represent monomers, counterions and crowders respectively. }
\begin{ruledtabular}
\begin{tabular}{cccc}
pair    & $\epsilon$ & $\sigma$ & $r_c$\\
\hline
 $m$ - $m$   & 1.0               & 1.0 & $2^{1/6}$\\ 
 $i$ - $i$   & 1.0               & 1.0 & $2^{1/6}$\\ 
 $m$ - $i$   & 1.0               & 1.0 & $2^{1/6}$\\ 
 $m$ - $c$   & $\epsilon_{mc}$   & 1.0 & $2.5$    \\ 
 $i$ - $c$   & 1.0               & 1.0 & $2^{1/6}$\\ 
 $c$ - $c$   & 1.0               & 1.0 & $2.5$    \\ 
\end{tabular}
\end{ruledtabular}
\end{table}

The PE, counterions,  and crowders are assumed to be in a medium of uniform relative dielectric constant $\varepsilon$. The electrostatic energy between any two charges $q_{1}$ and $q_{2}$ separated by distance $r$ is given by
\begin{equation}
U_{C}(r)= \frac{q_1 q_2}{4\pi\varepsilon\varepsilon_{0} r},
\label{eq.1}
\end{equation}
where $\varepsilon_{0}$ is the dielectric permittivity of vacuum.
The relative strength of electrostatic interactions among charged monomers along the PE chain is characterized by a dimensionless parameter $A$:
\begin{equation}
A=\frac{q^{2}\ell_B}{r_0},\quad \ell_B=\frac{e^2}{4\pi\varepsilon\varepsilon_{0} k_{B}T},
\label{eq.4}
\end{equation}
where $\ell_B$ is the Bjerrum length~\cite{khokhlov1994}  at which the electrostatic energy between two elementary charges $e$ is comparable  to the thermal energy $k_{B}T$. The strength of electrostatic interactions, $A$, can be tuned by either changing $q$ while keeping $\varepsilon$ fixed or by changing $\varepsilon$ while keeping $q$ fixed. In our simulations, we systematically scale the monomer charges $q$ to vary $A$. For $Z=1,2,3$, we do simulations for $A/A_c$ ranging from $0.001$ to $2.0$, where $A_c$ is the critical charge density for a PE in a good solvent in the absence of crowders. For each value of $A$, a range of values of $\epsilon_{mc}$ are explored. Lengths, temperatures and times are given in units of $\sigma$, $\epsilon /k_{B}$, and $\sqrt{m\sigma^{2}/\epsilon}$ respectively.

All the simulations are carried out using the molecular dynamics simulation package LAMMPS~\cite{plimpton1995fast} in NVT ensemble with periodic boundary conditions. The equations of motion are integrated using the velocity-Verlet algorithm~\cite{verlet1967computer, swope1982computer} with a time step of $0.001$. The system is initially equilibrated for $10^6$ steps and the analyses are done over subsequent $10^7$ steps.  The temperature is maintained through a Nos\'{e}-Hoover thermostat~\cite{nose1984unified, hoover1985canonical}. The long ranged Coulomb interactions are evaluated using the particle-particle/particle-mesh (PPPM) technique~\cite{hockney2021computer}.

\section{Results}
\subsection{\label{sec:results1}Collapse transition in PEs with crowder interactions}

\begin{figure}
\center
\includegraphics[width=\columnwidth]{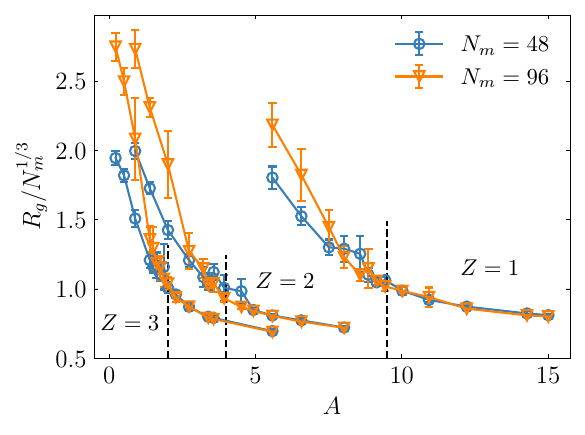}
\caption{The variation of radius of gyration, $R_g$, with charge density $A$ for two lengths of the PE, $N_m=48, 96$, in the absence of crowders. The collapsed phase is identified to be the regime when the data for $R_g/N_m ^{1/3}$ collapse on top of each other. The vertical dashed lines correspond to $2.0, 4.0, 9.5$ and are identified as approximate  values of $A_c$.}
\label{fig:critical_Ac}
\end{figure}

Before investigating the influence of attractive neutral crowders on the conformational behavior of charged PEs, we first determine the critical charge density, $A_c$, for the extended-collapsed transition of a single PE in good solvent conditions, in the absence of crowders.  To determine $A_c$, we monitor the radius of gyration, $R_g$, as a function of $A$, for all the three counterion valencies. $R_g$ is defined as
\begin{equation}
R_g ^2 = \frac{1}{2N_m ^2}\sum_{j=1}^{N_m} \sum_{k=1}^{N_m} |{\vec r}_j - {\vec r}_k|^2,
\label{eq:rg-def}
\end{equation}
where ${\vec r}_i$ denotes the position vector of the $i$-th monomer. The variation of $R_g$ with $A$  for two different values of $N_m$ is shown in Fig.~\ref{fig:critical_Ac}. By identifying the collapsed phase  to be the regime where the data for $R_g/N_m^{1/3}$ collapse onto one curve, we obtain $A_c \approx 9.5, 4.0, 2.0$ for $Z=1, 2, 3$ respectively. These values are larger than those obtained earlier for the PE in poor solvent conditions~\cite{varghese2011phase}.

Our aim is to identify the different phases and obtain the phase diagram of the PE in the presence of attractive crowders, particularly in the $A$--$\epsilon_{mc}$ phase space. Towards this end, we determine  $R_g$ and asphericity, $\alpha$, at different attractive strengths, $\epsilon_{mc}$,  for a range of values of $A/A_c$. Asphericity, $\alpha$, is defined as~\cite{tom2017aggregation, dima2004asymmetry, theodorou1985shape} 
\begin{equation}
\alpha =\frac{\lambda_1 \lambda_2+\lambda_2 \lambda_3+\lambda_3 \lambda_1}{(\lambda_1+\lambda_2+\lambda_3)^2},
\label{eq:alpha-def}
\end{equation}
where $\lambda_1, \lambda_2, \lambda_3$ are the eigenvalues of the moment of inertia tensor and the value of $\alpha$ varies between 0 (sphere) and 1 (rod).  

\begin{figure}
\includegraphics[width=\columnwidth]{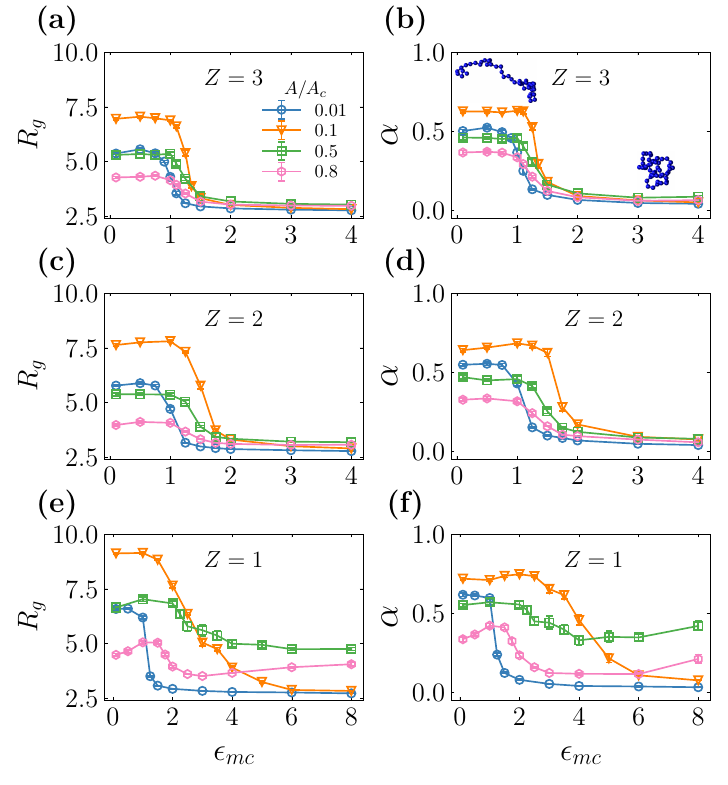}
\caption{$A/A_c < 1.0$: The radius of gyration, $R_g$, of the PE chain  and asphericity $\alpha$  as a function of  $\epsilon_{mc}$ for (a, b) $Z=3$, (c, d) $Z= 2$, (e, f) $Z = 1$, and $N_m = 48$.}
\label{fig:rgvsemcA<Ac}
\end{figure}

\begin{figure}
\center
\includegraphics[width=0.8\columnwidth]{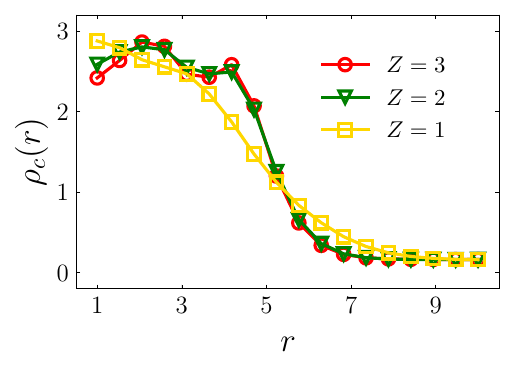}
\caption{$A/A_c=0.1$: Radial density function of crowders $\rho_{c}(r)$ measured from the center of mass of the collapsed PE chain for all the three valencies of counterions for $\epsilon_{mc} = 4.0$. }
\label{fig:rhocsmallA}
\end{figure}

We first focus on $A/A_c < 1$ (partial counterion condensation regime), where in the absence of crowders, the PE is in an extended conformation, the $CE$ phase, due to intra polymer  electrostatic repulsion. Consider the case of trivalent counterions, shown in Fig.~\ref{fig:rgvsemcA<Ac} (a). When the crowder strength $\epsilon_{mc}$ is relatively small, the value of $R_g$ closely resembles that observed without crowders, thus maintaining the extended conformation of the PE. However, as $\epsilon_{mc}$ increases further, a rapid reduction in $R_g$ is observed, indicating a pronounced collapse of the PE. Concurrently, the asphericity of the PE diminishes from a value close to $0.5$ to near zero, consistent with expectations for a collapsed phase, as depicted in Figure~\ref{fig:rgvsemcA<Ac} (b). To understand the origin of this attractive crowder induced collapse of PE,  we examine the radial density distribution of crowders, $\rho_c(r)$, measured from the center of mass of the PE in the collapsed phase ($\epsilon_{mc}=4.0$).  Fig.~\ref{fig:rhocsmallA} shows an excess crowder density within the collapsed PE region compared to the exterior, indicating that the collapse can be primarily attributed to bridging interactions with crowders rather than depletion effects. This specific collapsed phase, triggered by bridging interactions with neutral crowders, will be referred as the $CCB$ (bridging-induced collapsed) phase. Further evidence of the PE being collapsed for large $\epsilon_{mc}$  can be seen from the data collapse  of $R_g/N_m^{1/3}$, for two different PE chain lengths, in Fig. S1 (see Supp Info). We can thus identify a critical $\epsilon_{mc}^*$, for each $A/A_c<1$, beyond which an extended to collapsed transition ($CE$-$CCB$) is driven by bridging neutral crowders. This allows us to construct the $CE$-$CCB$ phase line in Sec.~\ref{sec:phase-diagram}. Similar trends are seen for divalent counterions as well [see Fig.~\ref{fig:rgvsemcA<Ac}(c),(d)]. 

The dependence of $R_g$ on $\epsilon_{mc}$ for monovalent counterions exhibits distinct characteristics compared to PEs with divalent and trivalent counterions, as shown in Fig.\ref{fig:rgvsemcA<Ac} (e). While the behavior of $R_g$ for monovalent counterions at $A/A_c=0.01$ and $0.1$ is similar to that of divalent and trivalent counterions, the transition occurs at larger critical values of $\epsilon_{mc}^*$. In contrast, for $A/A_c=0.5$ and $0.8$, the increase in $\epsilon_{mc}$ does not lead to a significant reduction in $R_g$. Moreover, the corresponding asphericity remains significantly different from zero even at large $\epsilon_{mc}$ values, as illustrated in Fig.\ref{fig:rgvsemcA<Ac} (f). Taken together, this suggests that the PE does not undergo collapse for intermediate $A$ values in the presence of monovalent counterions. We discuss the implications of this feature for the phase diagram for monovalent counterions in Sec.~\ref{sec:phase-diagram}.

\begin{figure}
\center
\includegraphics[width=\linewidth]{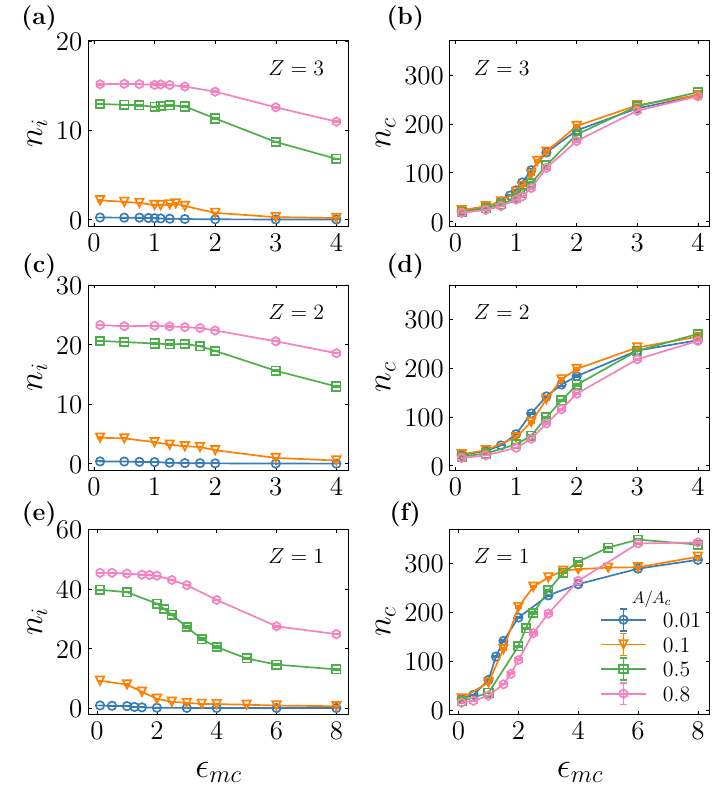}
\caption{The number of condensed counterions, $n_i$, and condensed crowders, $n_c$, within a distance $2\sigma$ of the PE chain as a function of $\epsilon_{mc}$ for $A/A_c < 1.0$ for (a, b) $Z=3$, (c, d) $Z= 2$, (e, f) $Z = 1$.}
\label{fig:nincvsemcAltAc}
\end{figure}

To understand the interplay between the condensed counterions and crowders as a function of increased monomer-crowder interaction strength, we calculate their numbers, $n_i$ and $n_c$ within a distance of $2\sigma$ of any monomer. In the regime of very low $A$ values, where Manning condensation has yet to take effect, $n_i$ remains close to zero across all $\epsilon_{mc}$ values, while $n_c$ exhibits an increase that eventually saturates at larger $\epsilon_{mc}$ values, for $Z=1,2,3$, as depicted in Fig.~\ref{fig:nincvsemcAltAc}. While the behavior of $n_c$ is largely unaffected by variations in $A/A_c$, a clear reduction in the number of condensed counterions with increasing $\epsilon_{mc}$ is observed for values of $A/A_c \lesssim 1$. A consequent increase in repulsive electrostatic energy, due to loss of charge renormalizing counterions, is however compensated by the increased attractive van der Waal energy due to neutral crowders, as can be seen from Fig.~S2, and drives the collapse of PE.

\begin{figure}
\center
\includegraphics[width=\linewidth]{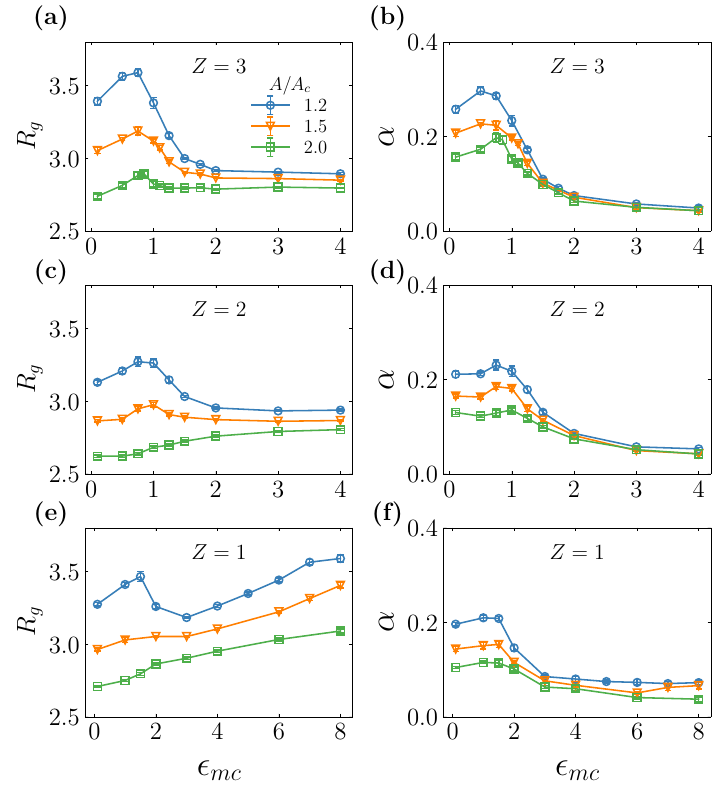}
\caption{$A/A_c > 1.0$: The radius of gyration, $R_g$, of the PE chain  and asphericity $\alpha$  as a function of  $\epsilon_{mc}$ for (a, b) $Z=3$, (c, d) $Z= 2$, (e, f) $Z = 1$, and $N_m = 48$.}
\label{fig:rgvsemcA>Ac}
\end{figure}

\begin{figure}
\center
\includegraphics[width=0.8\linewidth]{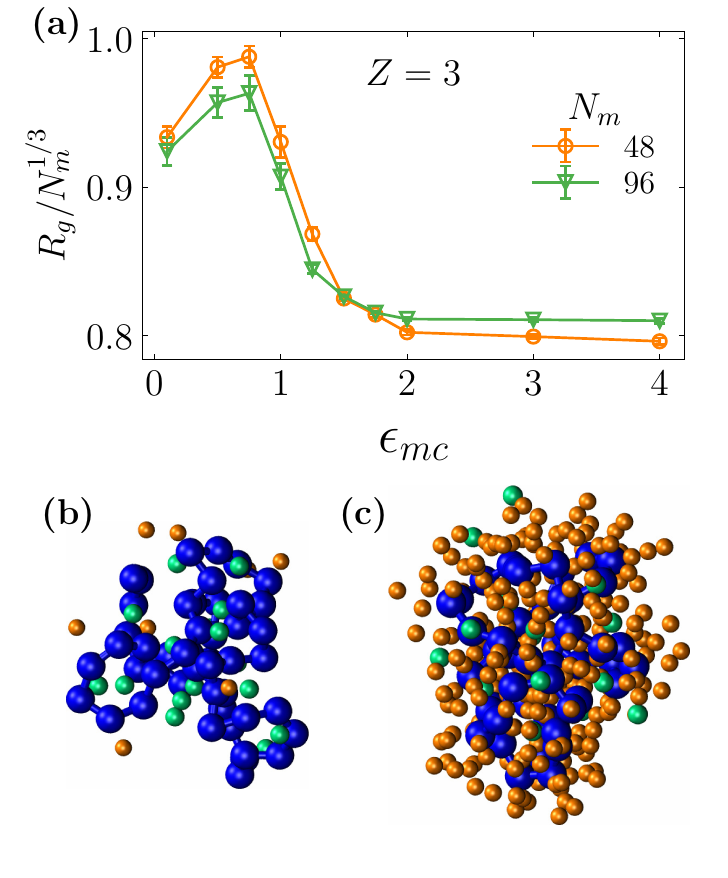}
\caption{(a) The scaled $R_g$ for two different $N_m=48,96$ for $A/A_c = 1.2$ and $Z=3$. (b-c) Snapshots of typical configurations for $N_m = 48$ and $Z=3$ with $A/A_c = 1.2$ for (b) $\epsilon_{mc} = 0.1$ and (c) $\epsilon_{mc}=3.0$. The monomers, counterions and crowders are colored by blue, lime green, and golden yellow respectively.}
\label{fig:gofr}
\end{figure}

We now focus on the results for $A/A_c >1$, where in the absence of crowders, complete counterion condensation occurs and the PE is in a collapsed conformation. Consider first the case for trivalent counterions. When $\epsilon_{mc}$ is increased from zero, an increase of $R_g$  can be seen from Fig.~\ref{fig:rgvsemcA>Ac}(a) suggesting swelling of the collapsed conformation. For this range of $\epsilon_{mc}$ values, the crowders effectively act as good solvent conditions and the swelling can be attributed to enhanced attractive interaction energy between PE and crowders. For larger values of $\epsilon_{mc}$, the $R_g$ saturates to a value that is independent of value of $\epsilon_{mc}$. Throughout this entire range of $\epsilon_{mc}$, $R_g$ consistently maintains values significantly smaller than those corresponding to the extended phase of the PE for $A/A_c < 1$ (as seen in Fig.\ref{fig:rgvsemcA<Ac}(a)), demonstrating that the PE retains a collapsed conformation. This is further supported by the consistently low values of the asphericity, $\alpha$, shown in Fig.\ref{fig:rgvsemcA>Ac}(b). 
In order to quantitatively validate the collapsed state of the PE across the entire $\epsilon_{mc}$ range, Fig.\ref{fig:gofr}(a) illustrates the data collapse of $R_g$ onto one curve for two distinct system sizes when scaled by $N_m^{1/3}$. Similar trends are observed for both divalent (see Fig.\ref{fig:rgvsemcA>Ac}(c-d)) and monovalent (see Fig.\ref{fig:rgvsemcA>Ac}(e-f)) counterions.

\begin{figure}
\center
\includegraphics[width=\linewidth]{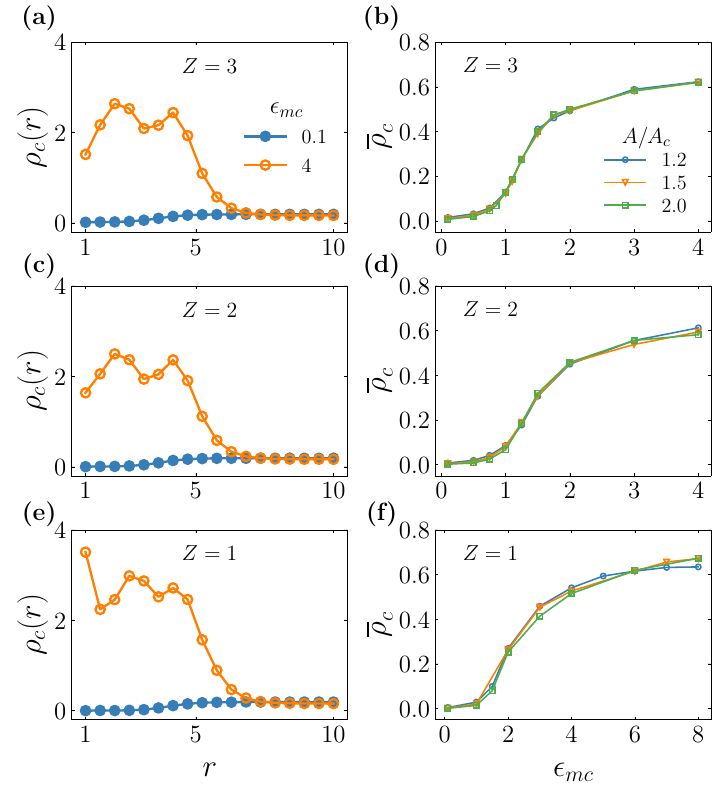}
\caption{Radial density function of crowders $\rho_{c}(r)$ measured from the center of mass of the collapsed PE chain for all the three valencies of counterions considered in the study for $\epsilon_{mc} = 0.1, 4$. The data shown in this figure is for system with crowder density $\rho_c = 0.05$ and $A/A_c = 2.0$. The critical Radial density function $\rho_c(r)$ of crowders within collapsed conformation for all the three valencies of counterions considered in the study as a function of $\epsilon_{mc}$. The transition from $CCI - CCB$ phase is identified \textit{via} this parameter. The data shown in this figure is for system with crowder density $\rho_c = 0.05$.}
\label{fig:rhocrowvsr}
\end{figure}

With the understanding that the PE consistently maintains a collapsed conformation across the entire range of $\epsilon_{mc}$ values, we now probe whether the nature of this collapsed phase remains uniform throughout. Snapshots of the PE(blue color particles), along with counterions(lime green particles) and crowders(golden yellow particles) within a distance of $2\sigma$ from the PE, are shown in Fig.\ref{fig:gofr}(b-c) for both small and large $\epsilon_{mc}$ values. In the case of small $\epsilon_{mc}$, the collapsed phase lacks crowders within the globule, forming what we refer to as the $CCI$ phase. In contrast, for large $\epsilon_{mc}$, a significant number of crowders are enclosed within the globule, defining the $CCB$ phase, as discussed previously. To differentiate between these two collapsed phases ($CCI$ and $CCB$), we analyze the radial density distribution of the crowders from the center of the collapsed globule, $\rho_c(r)$, as shown in Fig.\ref{fig:rhocrowvsr}. For all counterion valencies, it can be seen that the crowders are depleted from the center of the collapsed globule under small $\epsilon_{mc}$ conditions ($CCI$ phase), in contrast to the bulk crowder density ($\rho_c^{bulk}=0.05$). In contrast, for large $\epsilon_{mc}$ values ($CCB$ phase), a pronounced crowder overpopulation is observed within the globule ($\rho_c^{center}\approx 2.5$), strongly indicating that the two collapsed phases are distinct.

\begin{figure}
\center
\includegraphics[width=\linewidth]{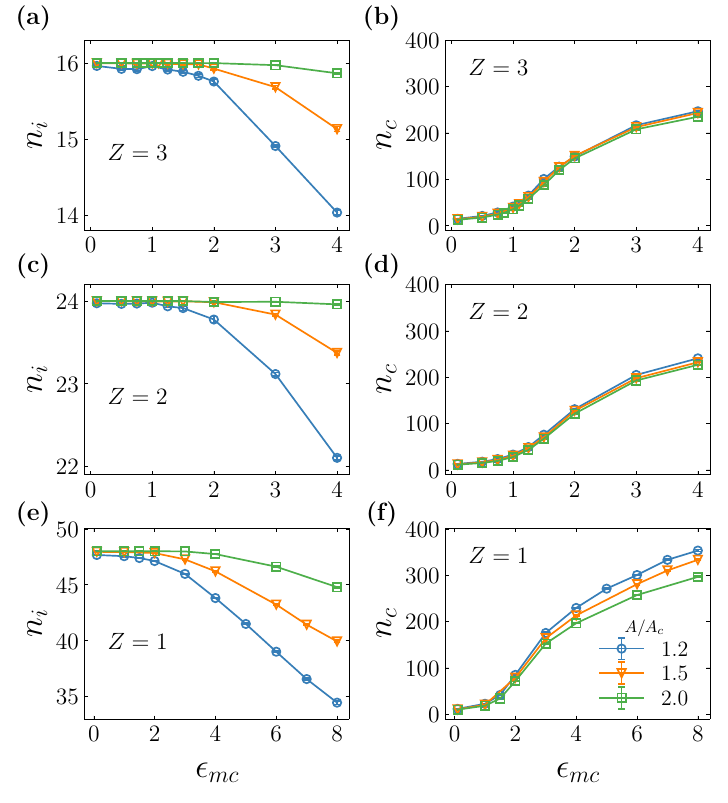}
\caption{The number of condensed counterions, $n_i$, and condensed crowders, $n_c$, within a distance $2\sigma$ of the PE chain as a function of $\epsilon_{mc}$ for $A/A_c > 1.0$ for (a, b) $Z=3$, (c, d) $Z= 2$, (e, f) $Z = 1$.}
\label{fig:nincvsemcAgtAc}
\end{figure}

To identify the transition between the two collapsed $CCI$ and $CCB$ phases, we define a mean density of crowders within the collapsed phase, $\overline{\rho}_c$, as the density within a radius $R_g/2$. As can be seen from the right panels of Fig.~\ref{fig:rhocrowvsr}, $\overline{\rho}_c$ remains close to zero for small $\epsilon_{mc}$ values across all counterion valencies, then sharply increases beyond a critical threshold $\epsilon_{mc}^*$. Remarkably, the $\overline{\rho}_c$ data exhibits minimal dependence on $A/A_c$, implying that the critical $\epsilon_{mc}^*$ is relatively independent of $A$ for $A/A_c>1$. We now ask whether the introduction of crowders leads to the displacement of condensed counterions from the $CCI$ phase, as a function of $\epsilon_{mc}$? For all counterion valencies, even though $A$ is large, as $\epsilon_{mc}$ is increased counterions are expelled from the PE globule, as can be seen from the variation of $n_i$ with $\epsilon_{mc}$ in Fig.~\ref{fig:nincvsemcAgtAc}. Correspondingly, the number of condensed crowders, $n_c$ increases sharply to compensate for the increased electrostatic repulsion (see right panels of Fig.~\ref{fig:nincvsemcAgtAc}). It is worth noting that the number of condensed crowders, $n_c$, exceeds that of the displaced counterions by an order of magnitude. The consequent increase in electrostatic energy is compensated by the attractive van der Waal energy as can be seen from Fig.~S3.

\subsection{Conformational phase diagram of PEs with crowder interactions \label{sec:phase-diagram}}
In Sec.~\ref{sec:results1}, we showed that PE can exist in three phases in the presence of neutral crowders. (1)$CE$: extended conformations for low charge density, below a critical value, and weak PE--crowder attractive interactions, (2)$CCB$: collapsed conformations for charged density values less than or near the critical value and for strong PE--crowder attractive interactions, and (3)$CCI$: collapsed conformations for charged density values higher than the critical value, primarily due to counterion condensation. We now explore the complete phase diagram of the system in $A$-$\epsilon_{mc}$ plane and demarcate the phase boundaries.

We choose two different parameters to locate the phase boundaries among the three phases $CE$-$CCB$ and $CCI$-$CCB$.  Similar to our previous work~\cite{garg2023conformational}, variation of $R_g$ with $\epsilon_{mc}$ curves for various values of $A/A_c$ will be used for $CE$-$CCB$ phase boundary as follows. A curve from $R_g$ vs $\epsilon_{mc}$ plot is chosen which represents a specific $A/A_c$ value. This curve is then fitted to a general hyperbolic tangent function ($R_g(\epsilon_{mc})$). The first derivative $dR_g/ d\epsilon_{mc}$ gives the rate of change of $R_g$ with respect to $\epsilon_{mc}$. Solving $dR_g/ d\epsilon_{mc}$ = 0 yields the value $\epsilon_{mc}^*$ for which the rate of change of $R_g$ with respect to $\epsilon_{mc}$ is maximum. The value of $\epsilon_{mc}^*$ indicates the location of phase boundary between two phases on either side of $\epsilon_{mc}$. This process is repeated for all the different value of $A/A_c$. 
However, for obtaining the $CCI$-$CCB$ phase boundary, $R_g$ is not a good parameter as it does not vary much across the two collapsed phases. On the other hand, the parameter $\overline\rho_c$ clearly distinguishes between the two collapsed phases, as can be seen from  Fig.~\ref{fig:rhocrowvsr} and we use it to identify the $CCI$-$CCB$ phase boundary. Similar to how we analyzed $R_g$ to determine the phase boundary between $CE$-$CCB$, we pinpoint the critical value of $\epsilon_{mc}^*$ for the $CCI$-$CCB$ phase transition by identifying the point at which the rate of change of $\overline\rho_c$ with respect to $\epsilon_{mc}$ is maximized. To determine the $CE$-$CCI$ phase boundary, we simulate the PE for $N_m=48, 96$, and identify the critical value of $A$ to be that value beyond which the data for $R_g/N_m^{1/3}$ collapse onto one curve.

\begin{figure}
\center
\includegraphics[width=\linewidth]{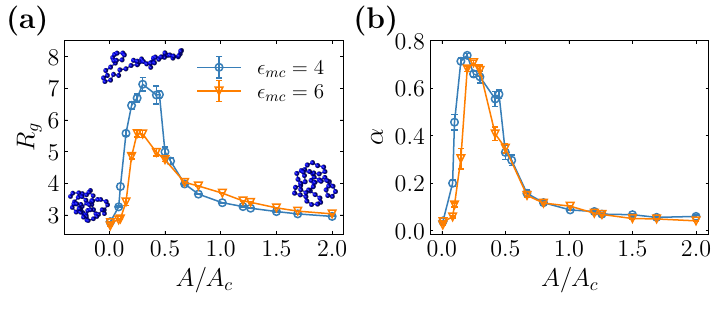}
\caption{$Z=1$: The variation of (a) $R_g$ and (b) $\alpha$ with $A/A_c$ for fixed $\epsilon_{mc}=4,6$. The snapshots in (a) correspond to $\epsilon_{mc}=4.0$.}
\label{fig:rgvsA}
\end{figure}

\begin{figure*}
\center
\includegraphics[width=\linewidth]{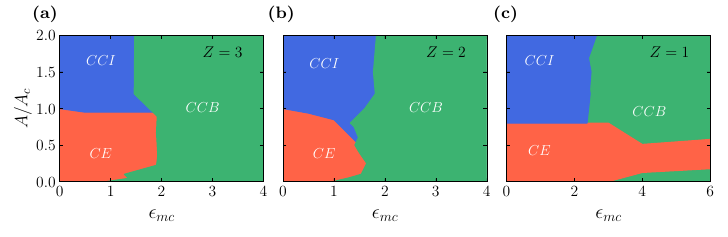}
\caption{The conformational phase-diagram of the PE in the $A/A_c$-$\epsilon_{mc}$ for (a) $Z=3$, (b) $Z=2$, and (c) $Z=1$.  The $CE$, $CCI$ and $CCB$ phases are represented in red, blue, and green colors respectively.}
\label{fig:PD_AllZ}
\end{figure*}

The phase diagram in the $A$-$\epsilon_{mc}$, thus obtained from the two analysis described above, can be seen in Fig.~\ref{fig:PD_AllZ}. For all the three counterion valencies, the three direct phase transitions $CE$-$CCI$, $CE$-$CCB$, and $CCI$-$CCB$ are possible.  However, for monovalent counterions, for the range of $\epsilon_{mc}$ considered, there is no $CE$-$CCB$ phase transition for intermediate values of $A$. This implies that, for fixed $\epsilon_{mc}\gtrsim 2$ and increasing $A/A_c$, we expect a reentrant $CCB$-$CE$-$CCB$ phase transition.  If this were the case, then if the system is simulated at a fixed large $\epsilon_{mc}$ and by increasing $A$, the PE should undergo collapsed-extended-collapsed transitions. The results for such simulations are shown in Fig.~\ref{fig:rgvsA}. For intermediate values of $A/A_c$, the PE is extended, as can be seen from the snapshot, with a corresponding large value of $R_g$ compared to small and large $A/A_c$ [see Fig.~\ref{fig:rgvsA}(a)]. For small and large $A/A_c$, the PE is collapsed as can be seen from both the snapshots [see Fig.~\ref{fig:rgvsA}(a)], as well as asphericity being close to zero [see Fig.~\ref{fig:rgvsA}(b)]. Thus we confirm that for intermediate $A/A_c$, attractive crowders are unable to drive an extended to collapsed transition.

We also observe that the critical value $\epsilon_{mc}^*$ for the $CCI$-$CCB$ phase transition is largely independent of $A$. This is surprising, as one would expect that for higher values of parameter $A$, the CCI phase's stability would strengthen, necessitating a correspondingly larger $\epsilon_{mc}$ to induce its destabilization. A discussion of the implications for the different theories of PE can be found in Sec.~\ref{sec:discussion}. Further, the $CE$-$CCI$ phase boundary is largely independent of $\epsilon_{mc}$ for small $\epsilon_{mc}$ and occurs around $A/A_c \approx 1$. This is to expected as the transition is driven by condensed counterions, and crowders do not play any significant role. For intermediate values of $\epsilon_{mc}$, the presence of crowders decreases $A_c$. 

\section{Discussion and Conclusion \label{sec:discussion}}

The notion of macromolecular crowding in cellular environments affecting the stability of biopolymers, such as proteins, is now well-known~\cite{minton1981excluded, speer2022macromolecular}. However, most of the studies on crowders treat the crowders as hard particles, whose primary mode of interaction with proteins or model polymers is via depletion: the entropically driven exclusion of repulsive crowders leads to the collapse of the polymer/protein. Only recently, the role of soft attractive interactions between polymers and crowders has been explored, and it has been implicated in various phenomena such as chromosome organization, LLPS (liquid-liquid phase separation), etc~\cite{erdel2018formation, king2021phase, ryu2021current, azaldegui2021emergence,ryu2021bridging, brackley2020polymer, brackley2013nonspecific, barbieri2012complexity, malhotra2021unfolding}. Simple polymer models have been effectively used to understand the role of soft attractive interactions between crowders and biopolymers like chromatin, demonstrating their impact on the conformational landscape. These studies have investigated the influence of both attractive polymer-crowder interaction energy and crowder density. However, it should be noted that most of these models are based on neutral polymers, and their relevance to inherently charged biopolymers remains unclear.

Like-charged polymers, polyelectrolytes(PE), have been shown to undergo a counterintuitive extended to collapsed transition, in the presence of oppositely charged counterions, when the charge density of the backbone, $A$, is larger than a critical value $A_c$~\cite{varghese2011phase,brilliantov98, tom2016mechanism, tom2017regimes, Kremer1993, Kremer1995,winkler98, brilliantov98, Melnikov1999, Holm, anoop11, Dobrynin2005, Chertovich2016, Chertovich2016a}.  This dependence of the PE conformations on $A$ suggests an analogy with monomer-monomer interaction energy ($\epsilon_{mm}$) in neutral polymers, where small and large $\epsilon_{mm}$ values correspond to extended and collapsed states, respectively.  Recently, we obtained a detailed phase diagram for neutral polymers in the presence of attractive crowders in the $\epsilon_{mm}$-$\epsilon_{mc}$ phase, where $\epsilon_{mc}$ represents the polymer-crowder interaction strength~\cite{garg2023conformational}. In this paper, we explored the conformations of PEs  in the $A$-$\epsilon_{mc}$ phase space.  We addressed the following questions: (1) Do neutral crowders differentially affect the conformations of PE, depending on the charge density of the PE ($A$)? (2) What is the role of the  valency ($Z$) of the counterions? (3) What is the conformational phase diagram of PE in the $A$-$\epsilon_{mc}$ phase space? (4) How do attractive crowders impact charged polymers compared to neutral polymers? 

We demonstrate that attractive neutral crowders have the ability to collapse a charged extended polymers, despite the inherent repulsive interactions arising from like charges. This collapse phenomenon occurs for all counterion valencies. We show the existence of three distinct phases in the $A-\epsilon_{mc}$ phase space: (1) An extended phase denoted as $CE$, observed for low charge densities and weak polymer-crowder attractive interactions. For monovalent counterions, the $CE$ also exists for strong polymer-crowder attractive interactions. (2) A collapsed phase labeled as $CCI$, observed for high charge densities and weak polymer-crowder attractive interactions,  primarily driven by the condensation of counterions within the polymer structure. (3) A collapsed phase denoted as $CCB$, observed for strong polymer-crowder attractive interactions, regardless of the charge density (at least for $Z=2, 3$),  facilitated by the crowders acting as bridges or crosslinks between the monomers. 

Interestingly, the critical value of $\epsilon_{mc}$ required to induce the collapse of charged polymers ($CE$-$CCB$ transition) with divalent and trivalent counterions is not significantly different from that observed for neutral polymers, as reported in our previous work~\cite{garg2023conformational}. However, we find that monovalent counterions exhibit distinct effects on the phases compared to multivalent counterions. In particular, for intermediate values of $A$, we find that that the PE does not collapse for even quite large values of $\epsilon_{mc}$. This may be due to steric effects which limits the number of crowders that can condense onto the PE. If this were true, then smaller sized crowders should induce a $CE$-$CCB$ transition for intermediate values of $A$, because a larger number of crowders can be accommodated around the PE. Studying the effects of crowder size on the phase diagram is a promising area for future study. 

Across the $CE$-$CCB$ transition, the number of crowders increases substantially compared to the decrease in the number of condensed counterions, strongly suggesting that a much larger number of neutral crowders is required to collapse the charged PE to overcome the strong electrostatic repulsion. This difference in number of non-polymer particles inside the collapsed phase of the PE can lead to new subsets of collapsed phases, including swollen phases and can potentially affect the conformational landscape. The significant increase in attractive van der Waals interaction energy between the polymer and crowders also supports this notion. 

We observe an intriguing phenomenon regarding the $CCI$-$CCB$ phase boundary, which appears to remain largely unaffected by changes in $A$. This outcome is unexpected, as one would intuitively anticipate that for higher values of parameter $A$, the CCI phase's stability would strengthen, necessitating a correspondingly larger $\epsilon_{mc}$ to induce its destabilization. If $A$ was indeed analogous to $\epsilon_{mm}$ of neutral polymers, then it might be anticipated that the critical value of $\epsilon_{mc}$ ($\epsilon_{mc}^*$) would exhibit a linear increase with $A$, as suggested by Garg et al.~\cite{garg2023conformational}. The independence of the $CCI$-$CCB$ phase boundary on the variations in parameter $A$, holds particular significance in the context of theories surrounding the collapse of charged polymers. Various competing hypotheses, such as the ionic solid theory, dipole theory, and counterion fluctuation theory, have been posited to explain the counterintuitive collapse or aggregation of similarly charged polyelectrolytes (PEs)~\cite{brilliantov98, tom2016mechanism, tom2017regimes, bloomfield1997dna, cherstvy2010collapse, gordievskaya2018interplay, muthukumar2004theory, solis2000collapse, Dobrynin2005, varghese2011phase, khokhlov1994, Kremer1993, Kremer1995, winkler98, BoJon99, Holm, Chertovich2016, manning1969limiting, Khokhlov1980, khokhlov1994, tom2017aggregation, anvy16, varghese2012aggregation, pincus98}.  In the dipole theory, the parameter $A$ serves to enhance the effective value of $\epsilon_{mm}$, thereby leading to the prediction of a linear relationship between $\epsilon_{mc}^*$ and $A$, which is in contradiction to our results.  Our previous molecular dynamics simulations corroborate the counterion fluctuation theory, which argues that the emergent attractive forces are due to density fluctuations of densely packed counterions moving freely within the PE globule~\cite{brilliantov98, tom2016mechanism, tom2017regimes, Chertovich2016}. How the inclusion of neutral crowders in the overall picture of PE with counterions affects the predictions of counterion fluctuation theory would be part of the future study. It is to be noted that, we previously showed the underlying mechanism that governs the aggregation of multiple similarly charged PEs when their charge densities exceed the critical threshold established for the collapse of an individual PE~\cite{anoop12, tom2017aggregation, anvy16}. Thus, the implications of understanding the role of neutral crowders for aggregation mechanism of multiple PEs would also be important to understand. 

Moreover, the Flory-Huggins theory furnishes predictions regarding the conformational behavior of neutral polymers, accounting for diverse entropic and energetic factors influencing the free energy dynamics~\cite{de1979scaling}. However, the introduction of charges to the polymer structure necessitates the integration of electrostatic contributions into the free energy, accomplished through a Debye-Huckel term~\cite{overbeek1957phase}. Our previous work~\cite{tom2016mechanism, tom2017regimes} has demonstrated the requirement for additional terms within the virial expansion to comprehensively address the electrostatic influence on the free energy. The valency of counterions has a significant impact on the attractive volume interactions within the PE, as it determines the number of condensed counterions inside the collapsed PE globule~\cite{tom2017regimes}. The introduction of attractive neutral crowders further complicates the scenario, as they compete with the condensed counterions, thereby enhancing the attractive interactions between the polymer and the crowders. These contrasting behaviors underscore the pivotal role played by the counterion valency in shaping the conformational landscape of charged polymers in the presence of neutral crowders. This can have profound implications for polymer models used for understanding phenomena such as chromatin organization and LLPS. Models such as strings and binders (SBS)~\cite{chiariello2016polymer,barbieri2012complexity} or stickers-and-spacers~\cite{choi2020physical} are simple neutral polymer models (lattice, mean field, etc.) used to describe chromatin organization, folding, or membraneless organelle formation via LLPS. These models primarily consider specific interactions with binding molecules, modeled as simple crowders. We expect significant changes to the understanding of these systems when the inherent charges of these biopolymers are incorporated, and the role of counterion valencies and the strength of soft attraction between biopolymers and crowders are modeled. Understanding how the introduction of charges in simple polymer models will affect the comparison with experimental Hi-C data  will be part of future work.

\section{Supplementary Material}
Additional Data included.

\begin{acknowledgments}
	The simulations were carried out on the high performance computing machines Nandadevi at the Institute of Mathematical Sciences.
\end{acknowledgments}


%

\end{document}